%% file: main.tex
\documentclass[letterpaper]{article} 
\usepackage[]{aaai2026}  
\usepackage{times}  
\usepackage{helvet}  
\usepackage{courier}  
\usepackage[hyphens]{url}  
\usepackage{graphicx} 
\usepackage{natbib}  
\usepackage{caption} 
\input{content/preamble}

\usepackage{newfloat}
\DeclareCaptionStyle{ruled}{labelfont=normalfont,labelsep=colon,strut=off} 
\title{\textit{The Essentials of AI for Life and Society}: \\ A Full-Scale AI Literacy Course Accessible to All}
\author{
  Zifan Xu,
  Kristen Procko,
  Michael Munje,\\
  Kristin Patterson,
  Lea Sabatini,
  Joydeep Biswas,
  Peter Stone
}
\affiliations{
    The University of Texas at Austin\\
    2317 Speedway, Stop D9500, Austin, TX 78712
}

\begin{document}

\maketitle

\begin{abstract}

In Fall 2023, we introduced a new AI Literacy class called \textit{The Essentials of AI for Life and Society} (CS 109), a one-credit, seminar course consisting mainly of guest lectures, which was open to the entire university, including students, staff, and faculty.  Building on its success and popularity, this paper describes our significant expansion of the course into a full-scale three-credit undergraduate course (CS 309), with an expanded emphasis on student engagement, interactivity, and ethics-related components. To knit together content from the guest lecturers, we implemented a flipped classroom. This model used weekly asynchronous learning modules---integrating pre-recorded expert lectures, collaborative readings, and ethical reflections---which were then unified by the course instructor during a live, interactive discussion session.  To maintain the broad accessibility of the material (no prerequisites), the course introduced substantive, non-programming homework assignments in which students applied AI concepts to grounded, real-world problems. This work culminated in a final project analyzing the ethical and societal implications of a chosen AI tool.  The redesigned course received overwhelmingly positive student feedback, highlighting its interactivity, coherence, and accessible and engaging assignments. This paper details the course's evolution, its pedagogical structure, and the lessons learned in developing a core AI literacy course.  All course materials are freely available\footnote{\url{https://www.cs.utexas.edu/~pstone/Courses/309fall24/}} for others to use and build upon.
\end{abstract}


\section{Introduction}
The surge of public interest in artificial intelligence (AI), prompted by powerful large-language-model-based tools such as ChatGPT \cite{chatgpt}, has created an urgent campus-wide demand for accessible, non-technical AI education. Most existing resources were either too technical or too high-level, failing to provide foundational knowledge on how these new tools work in the context of the full range of AI paradigms and their inherent limitations. Our initial response in Fall 2023 was to rapidly develop an AI literacy course, \textit{The Essentials of AI for Life and Society}, which was open to the entire university community \cite{cs109eaai2024}. This one-credit online seminar featured different guest lectures each week covering topics in their domains of expertise, ranging from AI fundamentals to societal implications. Although participants reported significant gains in AI literacy, this initial offering revealed key pedagogical shortcomings: the series of guest lecturers lacked a cohesive narrative, assessment was based primarily on attendance, and readings were often too difficult for a general audience. 

This paper introduces our significant expansion of the course in such a way that directly addresses these shortcomings.
This new three-credit iteration was designed for a broad undergraduate student audience, aiming to provide a cohesive learning experience while introducing the full range of core AI concepts as well as ethical frameworks for evaluating their societal impacts. A key structural change was the adoption of a flipped classroom model, organized around weekly cycles: students completed asynchronous modules by Tuesday and then engaged in a live, synchronous discussion session on Thursday. The weekly asynchronous modules consisted of: (1) segmented lecture videos from the original course, with interleaved multiple-choice questions to assess comprehension; (2) more accessible reading materials, which students discussed using Perusall \cite{bharath2021perusall}, a collaborative annotation tool; and (3) a reflective essay assignment that required students to apply ethical frameworks to real-world AI scenarios. Each flipped classroom module was followed by a live, synchronous class session led by the course professor, a change that provided continuity and fostered a sense of community.

To move beyond the participatory and attendance-based assessments that characterized the one-credit iteration, CS 309 introduced five major non-programming homework assignments and a final project. These assignments provided hands-on experience with core AI concepts, guiding students to solve problems related to their own daily lives by applying principles such as planning, probabilistic modeling, and machine learning. The course culminated in a final project in which each student researched an AI tool relevant to their interests and wrote a paper analyzing its ethical and societal implications.

Based on the data reported in this paper, these pedagogical changes proved highly successful, resulting in even more positive student feedback and deeper engagement than in the first iteration. The contributions of this paper are as follows. 
\begin{itemize}
    \item We present a detailed, replicable model for teaching an AI literacy course with a scalable, comprehensive curriculum. 
    \item We describe a suite of assignments designed to make abstract AI concepts accessible without relying on programming skills or any other prerequisites.
    \item We provide evidence of the effectiveness of a single-instructor, flipped-classroom model for teaching interdisciplinary AI topics to a broad audience.
\end{itemize}

\input{content/related.tex}
\input{content/course.tex}
\input{content/reactions.tex}
\input{content/lessons_learned}

\section{Conclusion}
The evolution of \textit{The Essentials of AI for Life and Society} from a one-credit seminar to a three-credit undergraduate course demonstrates a successful model for building an accessible and scalable AI literacy curriculum. By adopting a single-instructor, flipped classroom design—combining asynchronous modules with live, interactive sessions—we addressed the limitations of the initial seminar and created a cohesive narrative that unified diverse topics presented. The addition of substantive non-programming assignments and deeply integrated ethics components enabled students from all disciplines to meaningfully engage with core AI concepts and their societal implications. Positive student feedback affirms the effectiveness of this shift, underscoring the value of a structured, interactive, and continuous learning experience.

The course structure, assignments, and pedagogical strategies detailed in this paper offer a replicable and effective blueprint for other institutions aiming to establish broad-based AI literacy education. In the spirit of advancing AI education for all, all course materials, including lecture videos, non-programming assignments, and project outlines, are freely available online for educators to use and build upon. Future iterations may focus on refining the difficulty and length of the reading assignments and exploring new ways to foster community in a large, hybrid-format class.

\section*{Acknowledgments}
\small
This study was evaluated and considered exempt from the IRB of the University of Texas at Austin and was conducted in accordance with relevant ethical guidelines. This work was supported in part by the National Science Foundation (CAREER-2046955, NRT-2125858, FAIN-2019844), and UT Austin's Good Systems grand challenge. The authors thank Dr. Kenneth Fleischmann for contributing course material on AI ethics. We are also grateful to the following faculty for contributing video lectures to the course: Roberto Martín-Martín, Adam Klivans, Ray Mooney, Kristen Grauman, Luis Sentis, Greg Durrett, Don Fussell, Craig Watkins, Sherri Greenberg, and Scott Aaronson. Video editing support was provided by the Liberal Arts Instructional Technology Services team at UT Austin. Any opinions, findings, and conclusions expressed in this material are those of
the authors and do not necessarily reflect the views of the sponsors. Peter Stone serves as the Chief Scientist of Sony AI and receives financial compensation for that role. The terms of this
arrangement have been reviewed and approved by the University of Texas at Austin in accordance with its policy on objectivity in research.

\bibliography{references}

\end{document}

%% file: content/preamble.tex
\usepackage{booktabs}
\usepackage{multirow}
\usepackage[table]{xcolor} 

\newcount\Comments  
\newcommand{\zifan}[1]{{\ifnum\Comments=1\textcolor{blue}{[zifan: #1]}\fi}}
\newcommand{\commentp}[1]{{\ifnum\Comments=1\textcolor{red}{[Peter: #1]}\fi}}
\newcommand{\jb}[1]
{{\ifnum\Comments=1\textcolor{cyan}{[Joydeep: #1]}\fi}}

%% file: content/related.tex
\section{Related Work}
\paragraph{AI Literacy}
Recent years have seen a growing interest in AI literacy across educational contexts, reflecting the increasing societal impact of AI technologies. AI literacy, broadly defined, encompasses not only an understanding of what AI technologies are and how they work, but also the ability to critically evaluate their applications, limitations, and ethical implications~\cite{ng2021ai,ng2023review}. While many AI literacy initiatives have been developed for K–12 audiences~\cite{williams2023review}, efforts at the university level remain comparatively sparse and often target students in technical majors. \citet{kong2021evaluation} describe a seven-hour AI literacy course for university students from diverse disciplines in Hong Kong, emphasizing accessible explanations and cross-disciplinary examples. Other higher education AI literacy offerings have focused on integrating ethical considerations into technical AI courses, such as embedding socio-ethical analysis in robotics curricula \cite{vekhter2023responsible} or incorporating responsible AI modules into machine learning classes~\cite{ai_ethics_course_survey}.

\paragraph{AI in University Curricula}
Universities across the U.S. and abroad have increasingly introduced dedicated undergraduate and graduate degrees in AI, with at least 30 institutions now offering AI degrees (e.g., Carnegie Mellon University’s B.S. in Artificial Intelligence; University of Texas at Austin’s M.S. in Artificial Intelligence). These programs typically assume substantial technical preparation. By contrast, non-technical AI courses for broad undergraduate audiences remain relatively rare. Notable exceptions include ``AI for Future Presidents'' at Yale~\cite{candon2025artificial}, which, like our course, requires no programming or advanced mathematics and is open to all majors. Similarly, Stanford University’s ``The Social and Economic Impact of Artificial Intelligence'' focuses on the societal implications of AI without requiring technical prerequisites, while Harvard’s ``Intelligent Systems: Design and Ethical Challenges'' integrates foundational AI concepts with real-world ethical debates.

\paragraph{Non-Programming AI Assignments}
Prior work has explored non-programming AI assignments through K–12 initiatives like AI4K12~\citep{touretzky2019envisioning} and unplugged activities, which convey AI concepts via manipulatives and scenario analysis, as well as undergraduate use of tools such as Google’s Teachable Machine~\citep{GoogleTeachableMachine} and MIT’s Scratch AI extensions to let students train models and explore bias without coding~\citep{druga2019mlforkids}. Ethical engagement activities—such as role-playing policy debates~\citep{shapiro2021using} or analyzing AI in media~\citep{burton2015teaching}—are also common. However, these efforts are often short-term or narrowly focused. In contrast, our course offers a full-semester progression of non-coding assignments that span planning, probabilistic reasoning, and model interpretation, with structured ethical reflection embedded in each module. Complementing this landscape, there are several widely accessible public introductory AI courses, e.g., AI for Everyone~\cite{DL_AI_for_everyone}, Elements of AI~\cite{Elements_of_AI}, and Google AI Essentials~\cite{Google_AI_Essentials}. These courses similarly prioritize conceptual understanding over programming. However, they are primarily designed for internet-scale, open access audiences, whereas our CS 309 course is structured as a full-semester core undergraduate offering with openly accessible course materials.

%% file: content/course.tex
\section{Course Design}
This section presents the details of the redesigned three-credit undergraduate course, \emph{The Essentials of AI for Life and Society} (CS 309), and its evolution from the one-credit, all-audience seminar (CS 109) offered in Fall 2023. It begins with the course learning objectives, then details the structure of its asynchronous and synchronous components, and concludes with a discussion of assignments and evaluation methods.
\begin{figure}[hbt!]
    \centering
    \includegraphics[width=0.85\linewidth]{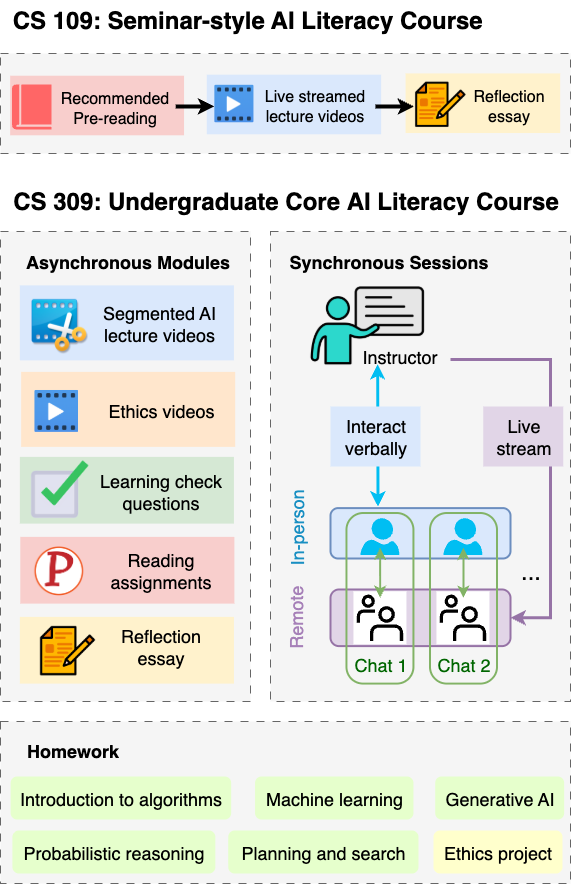}
    \caption{An overview of the course structure.}
    \label{fig:course_structure}
\end{figure}

\subsection{Curricular Objectives}
\begin{table*}[tbh!]
\centering
\resizebox{0.95\linewidth}{!}{
\begingroup
\rowcolors{2}{gray!10}{white}
\begin{tabular}{|l|l|l|l|}
\toprule
\textbf{Week} & \textbf{Technical and Societal Topics} & \textbf{Ethics Topics} & \textbf{Assignments and Projects} \\
\midrule
1 & Introduction & AI depictions in the media and summary of current ethical dilemmas& Introduction to Algorithms \\
2 & AI and Society & AI in the context of different ethical theories& Introduction to Algorithms \\
3 & Planning and Search & Human values and value-sensitive design& Planning and Search \\
4 & Intelligent Robotics & Codes of ethics& — \\
5 & Probabilistic Modeling & The global landscape of AI ethics guidelines& Probabilistic Reasoning \\
6 & Computer Vision & — & — \\
7 & Machine Learning Fundamentals & — & Machine Learning \\
8 & Machine Learning Paradigms & — & Machine Learning \\
9 & NLP (Large Language Models) & — & Generative AI \\
10 & Bias and Fairness in AI Models & — & — \\
11 & AI and Mis/disinformation & — & — \\
12 & Workplace Impacts, Economics, and Policy & — & — \\
13 & AI Alignment and Existential Threats & — & — \\
14 & Computational Foundations and Future Directions & — & Ethics Project \\
\bottomrule
\end{tabular}
\endgroup
}
\caption{Weekly schedule of asynchronous modules, showing core topics, integrated ethics themes, and associated assignments.}
\label{tab:schedule}
\end{table*}
CS 309 was designed to cultivate AI literacy, enabling students to engage with artificial intelligence in academic, professional, and civic contexts. By the end of the semester, students were expected to:
\begin{itemize}
\item Define AI in scientifically grounded terms and distinguish it from portrayals in science fiction;
\item Describe AI as an interdisciplinary field connecting computer science with the natural sciences, engineering, humanities, and social sciences;
\item Explain foundational AI concepts and relate them to applications in areas such as computer vision, robotics, natural language processing, and large language models;
\item Evaluate the benefits, risks, societal impacts, and ethical dimensions of AI technologies.
\end{itemize}
The idea of agency, of an \emph{autonomous agent} that can sense, plan, and act autonomously in the world, is a concept central to Artificial Intelligence, and is emphasized throughout the course. Starting from the working definition of Artificial Intelligence, to application topics including robotics, machine learning, and natural language, the course exposes students to the challenges of how to perceive the world robustly and infer the state relevant to its goals, how to plan a sequence of actions to move the world state from its current configuration eventually to the goal configuration, and how to translate such plans into individual actions, including motor actions, interactions with other agents, and information-gathering actions.

As summarized in Table~\ref{tab:schedule}, the curriculum spans foundational AI concepts (Weeks 1–3, 5, 7, 8),  domain-specific applications (Weeks 4, 6, 9), and broader societal and policy-oriented issues (Weeks 10-13), including bias and fairness, misinformation, and economic impacts. The course also fulfilled a university ethics requirement by asking students to apply ethical reasoning to real-world scenarios in both personal and professional contexts.

\subsection{Course Structure}
The course was redesigned into a blended format where asynchronous modules on the Canvas learning management system prepared students for the weekly live session. An overview of such a blended format is summarized in Figure \ref{fig:course_structure}. These modules were built from the seminars of the prior one-credit course, now segmented into shorter videos accompanied by multiple-choice questions to check for understanding. A major improvement was the revision of reading assignments: technically challenging articles were replaced with more accessible materials, which students collaboratively annotated and discussed in Perusall \citep{bharath2021perusall}. Foundational and application-focused modules were paired with substantive, non-programming homework assignments, and the semester culminated in a final project where students applied AI concepts to a real-world context and evaluated the societal and ethical implications of a chosen AI tool.

Ethics was deeply integrated to fulfill a university-wide requirement. Curated video excerpts from a dedicated ethics course were introduced in the week 1-5 modules. The  lectures began with an overview of current ethical dilemmas related to AI, connected these to varied principles of 10 major ethical theories, and progressed to consider the priorities emphasized in AI ethical standards developed across the globe.  Through weekly written reflections on complex AI application scenarios (e.g., the public deployment of AI-powered surveillance), students connected theories of ethics with real-world considerations to help shape their perspectives on AI technologies that are commonly used today, as well as deployment of new AI technologies in the future.

The weekly synchronous session, led by a single instructor, provided continuity across topics and served as a forum for live, interactive discussion—an intentional design choice responding to CS 109 feedback that rotating guest lecturers lacked cohesion. The instructor presented slides that incorporated representative student comments to present distinct viewpoints, using them to prompt class-wide discussion. 

The sessions were livestreamed from a professional studio, enabling simultaneous participation from both in-person and remote students. Each week, remote students participated via a moderated, text-based group chat while approximately 20 rotating “in-studio” students served as spokespeople for their remote groups, engaging verbally with the instructor. This hybrid arrangement combined the accessibility of online participation with the immediacy of in-person interaction, giving every student at least one opportunity for face-to-face engagement. Student feedback highlighted the effectiveness of this model, with one participant noting, “\textit{It really feels like we are all in this discussion together.}”

\section{Assignments and Projects}
\begin{figure*}
    \centering
    \includegraphics[width=\linewidth]{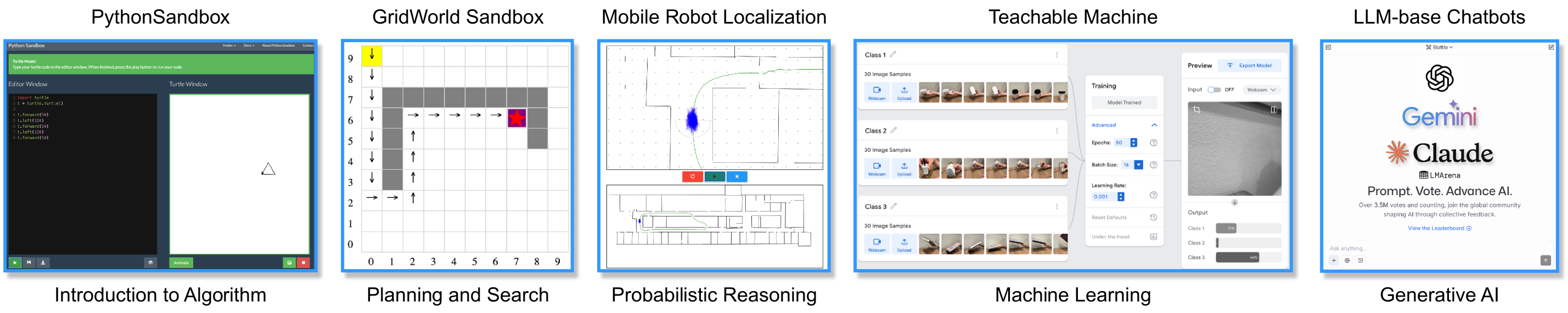}
    \caption{An overview of the non-programming tools used by five assignments.}
    \label{fig:tools}
\end{figure*}
This section describes the design of the five non-programming assignments, their learning objectives, and the tools students used to explore AI concepts in applied contexts. All assignments were developed to accommodate students with minimal levels of technical preparation, while still emphasizing a deep conceptual grasp of core AI techniques. An overview of the non-programming tools is shown in Fig. \ref{fig:tools}.

\paragraph{Introduction to Algorithms} The first assignment introduced students to the concept of algorithms as systematic, step-by-step procedures for solving problems—a foundational idea in artificial intelligence. Students selected a routine from their daily life (e.g., preparing a meal) and decomposed it into discrete steps, expressed in pseudocode, to encourage algorithmic thinking. They then implemented their algorithms using Python Sandbox~\cite{pythonsandbox} to control a virtual Turtlebot, programming it to move in desired patterns. Through this activity, students were introduced to core programming concepts such as syntax, debugging, and iteration, gaining hands-on experience in translating abstract procedures into executable instructions.

\paragraph{Planning and Search} In this assignment, students first learned the fundamentals of a planning domain, including its states, actions, and transitions. Using an interactive GridWorld sandbox (developed by the instructors of this course), they defined a planning problem with a start and goal state and manually found a solution: a valid sequence of actions to navigate an agent to the goal. The assignment then introduced Breadth-First Search (BFS), a classic algorithm that automatically solves these problems by exploring all possible states layer-by-layer to find the shortest path.

\paragraph{Probabilistic Reasoning} The third assignment demonstrated how AI systems manage uncertainty in the real world. Students began by identifying random events from daily life and distinguishing between independent and dependent events to build intuition for probability. These concepts were then applied to mobile robot localization, where a robot must estimate its position despite imperfect movements and noisy sensor data. Using an interactive simulator \cite{BiswasZhang2025ParticleFilters}, students observed how a LiDAR sensor perceives the environment and how measurement noise can lead to errors. The assignment concluded with an exploration of particle filter localization \cite{fox2001particle}, using a parameterized demo to visualize how particle samples represent and update uncertainty under varying levels of sensor and motion noise.

\paragraph{Machine Learning} The fourth assignment provided a hands-on introduction to machine learning using Google's Teachable Machine \cite{GoogleTeachableMachine}, a no-code platform. Students began by building their own image classification model, using a webcam to create a unique training dataset of nearby objects. They then tested their model on new images from the internet to experience the generalization gap—the model's performance drop on unseen data—and learned to close this gap by augmenting their training set with more diverse examples. The assignment concluded with an introduction to formal evaluation metrics, as students generated confusion matrices to compute accuracy, precision, and recall, and analyzed how class imbalance can yield deceptively high accuracy while masking poor performance on underrepresented categories.

\paragraph{Generative AI} The final assignment delved into Generative AI, focusing on the core techniques, capabilities, and limitations of Large Language Models (LLMs). Students first explored the mechanics of how LLMs work, including tokenization, probabilistic text generation, and decoding strategies. Using an interactive Chatbot Arena \cite{chiang2024chatbot}, they manipulated parameters like temperature and Top-p sampling to observe how these settings control the balance between creativity and coherence in the model’s output. Next, they benchmarked leading LLMs (e.g., ChatGPT-4o, Claude Sonnet) on tasks requiring mathematical, causal, and factual reasoning to compare their performance in a practical setting. The assignment concluded by examining key challenges like hallucination, where models produce incorrect information, and introduced the future of multimodal AI with a hands-on visual question answering task.

\paragraph{Ethics Project} The course culminated in a final Ethics Project, where students wrote a 3–5 page investigative report on an AI technology relevant to their personal or professional interests. To complete the report, students synthesized at least six sources to deconstruct their chosen technology (e.g., course readings, a peer-reviewed research paper, and articles in popular media). The analysis required them to identify the technology's core AI foundations, evaluate its strengths and societal risks, and perform an in-depth ethical assessment of one significant risk using the ethical frameworks introduced in the course. The project concluded with students proposing both technical improvements and concrete strategies to mitigate the identified ethical challenges.

%% file: content/reactions.tex
\section{Student Reactions}

\begin{figure*}[tbh!]
  \centering
  \includegraphics[width=\textwidth]{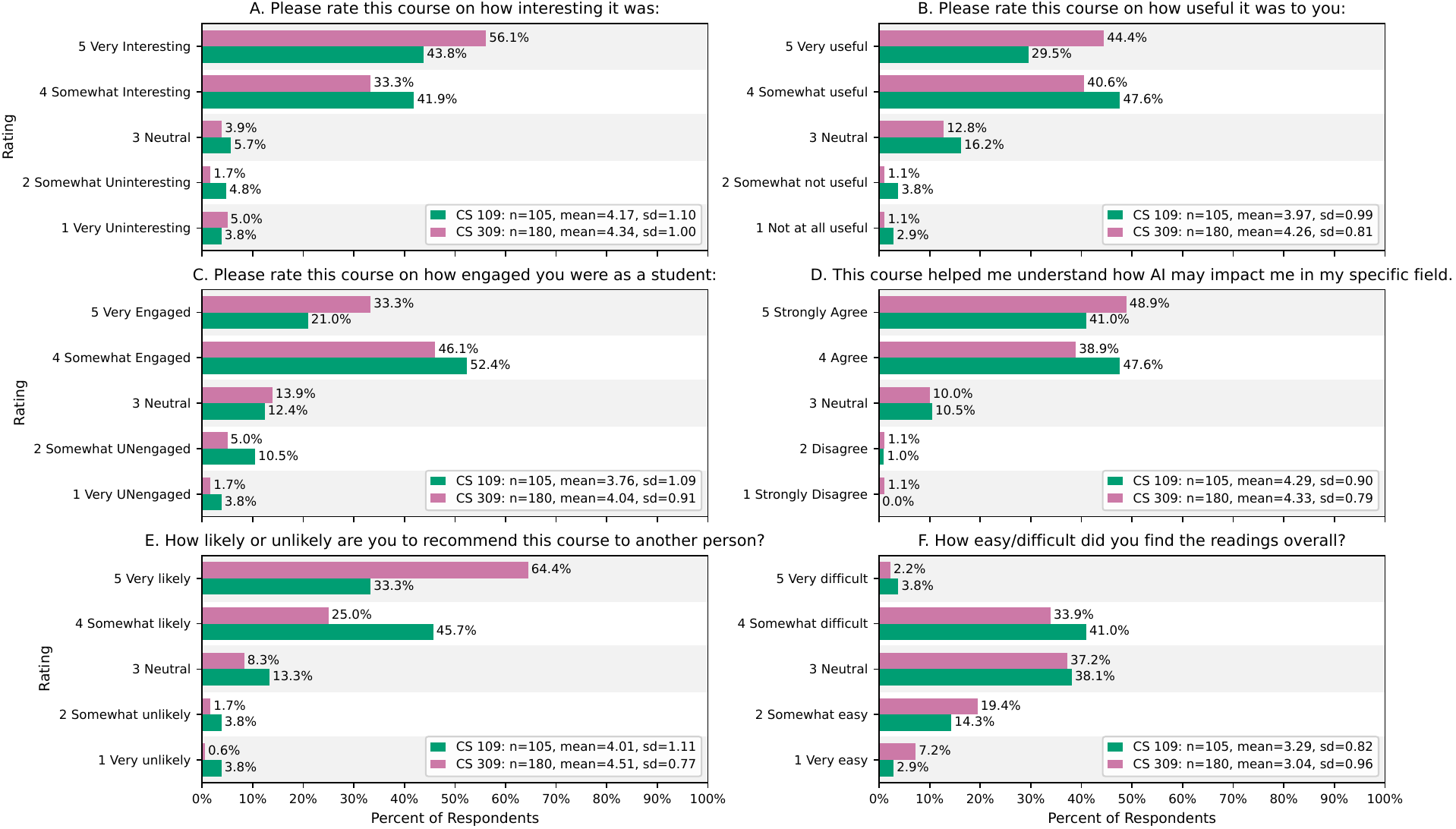}
  \caption{Course Polls comparing \textbf{CS 109} and \textbf{CS 309}.
  Bars show the percentage of respondents for each Likert option (per course).
  Panel legends display per-course mean and standard deviation.}
  \label{fig:course-polls}
\end{figure*}

\label{sec:student-reactions}

\noindent
To understand the effectiveness of CS 309 and identify areas for improvement, we conducted a post-course survey combining quantitative Likert-scale ratings with qualitative open-ended responses. This approach allows us to capture both broad trends in student satisfaction and more nuanced feedback on specific course elements. In addition to assessing the perceived importance of individual components such as lectures, assignments, and interactive activities, we examined thematic patterns in students’ comments. We also compared these results to the previous one-credit version of the course (CS 109) to evaluate the impact of the curricular changes implemented in the three-credit format. The survey reflects responses from 206 students who completed the course, with only 7 students withdrawing, resulting in a 96.7\% completion rate—indicating sustained engagement and an appropriate level of course difficulty that remained accessible to students from diverse academic backgrounds.

\subsection{Assignments: Quantitative Ratings}
At the conclusion of the semester, students rated the importance of each course component on a five-point Likert scale (Table~\ref{tab:feedback}). Lecture-based content, including asynchronous pre-recorded lectures and the live Thursday sessions, received the highest average ratings (above~4). Major assignments, including the five non-programming projects and the final ethics project, were also rated highly, reflecting  that the balance of conceptual and applied learning was valued by students.
\begin{table}[tbh!]
\centering
\resizebox{\linewidth}{!}{
\begin{tabular}{lcccccc}
\toprule
\textbf{Component} & \textbf{VU (1)} & \textbf{SU (2)} & \textbf{N (3)} & \textbf{SI (4)} & \textbf{VI (5)} & \textbf{Avg.} \\
\midrule
\multicolumn{7}{c}{\textbf{Asynchronous Module Components}} \\
\midrule
Lecture videos on AI concepts & 3 & 7  & 22 & 75 & 72 & 4.15 \\
Lecture videos on ethics      & 3 & 8  & 24 & 74 & 70 & 4.12 \\
Reading assignments        & 5 & 11 & 44 & 70 & 49 & 3.82 \\
Learning check questions & 7 & 16 & 41 & 78 & 37 & 3.68 \\
Reflection essays & 5 & 22 & 41 & 63 & 48 & 3.71 \\
\midrule
\multicolumn{7}{c}{\textbf{Synchronous Lecture Components}} \\
\midrule
Synchronous lecture & 3 & 3  & 18 & 56 & 99 & 4.37 \\
In-class group chat     & 6 & 18 & 46 & 59 & 50 & 3.72 \\
\midrule
\multicolumn{7}{c}{\textbf{Course Assignments and Projects}} \\
\midrule
Five assignments       & 5 & 11 & 37 & 56 & 70 & 3.98 \\
Ethics project                 & 4 & 9  & 42 & 62 & 62 & 3.94 \\
\bottomrule
\end{tabular}
}
\caption{Student ratings on the importance of specific course components (VU = very unimportant, SU = somewhat unimportant, N = neutral, SI = somewhat important, VI = very important; $n=181$).}
\label{tab:feedback}
\end{table}

\subsection{Assignments: Qualitative Themes}
Open-ended responses provided additional context. Three themes emerged:

\paragraph{Theme 1: Interactivity in CS~309.}
Students highlighted interactive elements such as Perusall discussions and Thursday activities as strengths. About 30\% of respondents mentioned Thursday components as a best aspect, emphasizing active discussion and incorporation of student input. Representative remarks included: ``The fact that student discussion is a major part of the course, even though it's asynchronous, is really nice,'' and ``Every class had multiple topics to workshop, creating original thought.''

\paragraph{Theme 2: Readings valued, with level and length as common caveats.}
More than one-quarter of responses identified the readings as helpful (e.g., ``I thought the readings were particularly useful---they were very well selected''), while also noting that some were long or technical. Students suggested refining selections to better support a wide range of backgrounds and adjusting Perusall expectations to prioritize comment quality over quantity.

\paragraph{Theme 3: Desire for stronger connections across components.}
Some students asked for clearer alignment among readings, prerecorded videos, Thursday activities, and weekly assignments. Comments noted that certain tasks felt disconnected from that week’s materials. Clarifying the ``through line'' each week (e.g., brief ``bridge'' slides, a short overview linking materials to in-class goals, or an alignment note on assignments) could make these connections more transparent.

\subsection{In-Studio Attendance}
CS~309 included a one-time in-studio requirement, meaning that students needed to attend (at least) one of the Thursday class sessions in person (as opposed to watching the live stream remotely). Likert responses indicated that students generally valued the experience (mean = 3.86) and some mentioned appreciating the in-class chat. As one student commented: “When you had a good group for Thursday in-class discussions it was useful to hear all of the different points of view. This helped me see things in a way I didn't originally think of.”  Students did not strongly prefer additional synchronous meetings (mean = 3.37). Open-ended feedback was mixed: some appreciated the opportunity to engage in person, while others emphasized the importance of preserving online accessibility. Several comments suggested small usability improvements to the live chat (e.g., threading or reactions) to make participation easier to follow.

\subsection{Comparisons to the Preliminary Offering}
We compared CS~309 to the prior one-credit seminar (CS~109). Figure~\ref{fig:course-polls} summarizes six poll items as distributions with per-panel means and standard deviations. Overall, CS~309 trends more favorably: higher ratings for interest, usefulness, engagement, perceived domain impact of AI, and likelihood to recommend, and lower perceived difficulty of readings.

\subsection{Summary}
Overall, students responded positively to CS~309. Lecture quality and interactive elements were clear strengths, and major assignments were viewed as valuable. The most common suggestions centered on calibrating reading level/length, refining Perusall expectations, and making weekly connections among materials and activities more explicit. These adjustments can build on what worked while improving clarity and consistency across the course.

%% file: content/lessons_learned.tex
\section{Lessons Learned}
\paragraph{Lectures.}
Many students praised the clarity and production quality of the lectures, which were among the highest-rated components (Avg.\ 4.37 for the synchronous session; 4.12–4.15 for prerecorded videos; see Table~\ref{tab:feedback}). However, several students wanted more ways to bring questions from the asynchronous materials into Thursday and to make links to the week’s readings more explicit.

\paragraph{In-class chats.}
Many students appreciated the interactivity of the in-class chats. However, the quality of the experience was directly tied to the group interactivity, which was variable due to automatic, random group creation each week. Though not possible with our learning management system, it would be beneficial to create groups of mixed background and levels of comfort with online chats, potentially using a survey at the beginning of the term. This may foster a productive discussion for more students.

\paragraph{Homework.}
Many students valued the major assignments and the ethics project (Avg.\ 3.98 and 3.94; Table~\ref{tab:feedback}). The most common suggestions focused on workload and week-to-week alignment (e.g., “Some of the five homework assignments were very time intensive,” “Homeworks seemed a lot different from what we learn during lectures”). To make the assignments more consistently effective, future iterations could clarify rubrics and expectations, more closely align tasks with the week’s materials, and better calibrate workload across weeks.